# ATLAS: a Cassegrain spectrograph based on volume phase holographic gratings

J.G. Robertson[a,b], K. Taylor[a], I.K. Baldry[a], P.R. Gillingham[a] and S.C. Barden[c]

[a] Anglo-Australian Observatory, PO Box 296, Epping, NSW 1710, Australia
[b] School of Physics, University of Sydney, NSW 2006, Australia
[c] National Optical Astronomical Observatories, PO Box 26732, Tucson, AZ 85726

## ABSTRACT

We are proposing a new spectrograph (ATLAS) which would revolutionise intermediate-dispersion observations at the AAT. Based on the new technology of Volume Phase Holographic gratings, and using transmission optics, ATLAS offers high throughput and a wide field (24′). It will be ideally suited to extensive surveys of faint objects. It has been designed with a collimated beam diameter of 150 mm, giving resolution $\lambda/\delta\lambda$ up to nearly 10,000 with a 1.5 arcsecond slit and good efficiency. It will be a dual-beam instrument, to maximise observing speed and allow optimised optical coatings to be used. The project is working towards its Concept Design Review which will occur during 2000.

**Keywords:** spectrograph, volume phase holographic, optical, survey spectroscopy

## 1. INTRODUCTION

Cassegrain intermediate dispersion spectroscopy has been a mainstay of activities at the Anglo-Australian Telescope (AAT) for many years, primarily using the RGO Spectrograph[1]. However, this instrument is no longer competitive, in either throughput or the wavelength range of a single exposure. The AAO is now planning a new Cassegrain spectrograph, **ATLAS** (**A**AO **T**unable **L**ittrow **A**rticulated **S**pectrograph), which will take advantage of the new Volume Phase Holographic (VPH) diffraction gratings. The combination of high efficiency VPH gratings, coated transmission optics, the large detector arrays now available, and the wide field at the AAT's Ritchey-Chrétien focus will allow ATLAS to do much more than simply update the RGO Spectrograph. ATLAS will have a field size of 24′, providing a powerful multi-object mode for extensive survey spectroscopy of faint objects. In this respect it will also replace the existing Low Dispersion Survey Spectrograph, and will complement the AAO's Two-degree Field (2dF) instrument (which performs survey spectroscopy of brighter objects over a still larger field). Wide-field survey work on 4-m class telescopes can be expected to play an important role in identifying objects suitable for later follow-up at the 8-m telescopes, in particular Gemini South and the VLT, in the case of surveys made at the AAT. There will be follow-up studies of fields from the Sloan and Chandra/XMM surveys, as well as stand-alone optical surveys.

As a high-throughput general purpose intermediate dispersion facility, ATLAS has a strong and very diverse scientific case. In addition to survey spectroscopy of objects to $V \sim 24$, it will offer single-object spectroscopy at resolutions $\lambda/\delta\lambda$ from 1,000 to about 10,000, making it well suited to a wide variety of studies. In order to make observations as efficient as possible, ATLAS will be a dual beam instrument. The preliminary specifications of ATLAS are given in Table 1, and an outline of the optical system is shown in Figure 1.

## 2. VOLUME PHASE HOLOGRAPHIC GRATINGS IN ATLAS

Volume Phase Holographic (VPH) gratings consist of a transparent layer of dichromated gelatin (DCG) in which the refractive index is modulated to form the lines of the grating. The principles of VPH gratings as applied to astronomical spectrographs have been given by Barden *et al*[2]. Here we mention the main points only, and concentrate on the differences in performance between VPH and surface reflection gratings in an astronomical spectrograph.

The thickness of the DCG layer is typically between 3 µm and 20 µm. It is because this thickness is greater than the wavelength of light that the gratings are referred to as *volume* phase holographic. This has the important result that the efficiency of the grating is high only in the direction satisfying the Bragg condition, which is equivalent to specular reflection from the planes of constant refractive index. Hence a general-purpose VPH spectrograph must allow a wide range of collimator – grating – camera angles, which we refer to as the articulated camera concept. The particular wavelength which satisfies the Bragg condition can be varied by tilting the grating with respect to the incident beam, effectively tuning the peak efficiency of the grating to a desired wavelength.

**Table 1** Principal properties of ATLAS

| | |
|---|---|
| Focal station | Cassegrain $f/7.9$ (primary mirror 3.89 m) |
| Spatial field diameter | 24 arcmin (215 mm) with $f/2.2$ camera |
| Collimated beam diameter | 150 mm |
| Optical system | transmission optics, VPH gratings at Littrow |
| Wavelength range | 360 – 550 nm (blue arm) |
| | 550 – 1000 nm (red arm) |
| Camera focal ratio | 1.7 – 2.2 (to be decided) |
| Detector arrays | 4096 × 4096 pixels |
| Detector pixel size | 15 μm |
| Image scale | 0.36 arcsec/pixel at $f/2.2$ |
| Spectral resolution (1.5″ slit) | 1,000 to 9,500 |
| Collimator-camera articulation angle | 0 – 90° |
| Optical throughput (target) | 90% (optics only; sol-gel + $MgF_2$ coatings) |
| Peak overall throughput | 40% (atmosphere, telescope, optics, grating, CCD) at 550 nm |
| Main modes | - multi-object spectroscopy using masks |
| | - longslit spectroscopy |
| | - direct imaging |
| Additional modes | - IFU for area spectroscopy over ~ 20″ × 20″ at R ~ 19,000 with $f/2.2$ camera |
| | - spectropolarimetry |
| | - cross-dispersed multi-order spectroscopy (?) |

It is this process which replaces the concept of blaze for a conventional surface grating: for VPH there is no fixed 'blaze wavelength'. A VPH grating does, however, have a design wavelength at which its performance is optimised. This is the wavelength at which the product of the thickness and the amplitude of refractive index modulations gives full modulation of the phase of the wavefront for the appropriate angle of incidence. When tuned to this wavelength, the grating's peak efficiency will take its highest value. When tuned to other wavelengths the peak efficiency will be somewhat lower. The envelope of the peak efficiencies as a grating is tuned is referred to as the 'superblaze' curve. Figure 2 illustrates that VPH gratings may be tuned by changing the angle of incidence, as well as showing their high peak efficiency.

The advantages of VPH gratings are:

1. High efficiency: throughput greater than 80% to 90% can be achieved at the peak of first order. This is significantly better than the 60% - 70% typically obtainable from conventional ruled surface reflection gratings.

2. As transmission gratings, they can be used in Littrow configuration (i.e. the incident and diffracted beams make equal angles with the grating normal). This eliminates beam dilation and the consequent increase in camera aperture. Transmission optics also avoids vignetting by central obstructions.

3. Use in transmission also allows shorter pupil relief between the grating and both the collimator and camera. This again reduces the size of the required camera aperture, and increases the field of view.

4. Gratings are available with line densities from about 100 lines/mm to 6000 l/mm. Thus higher dispersions can be obtained in first order than is the case for ruled surface gratings, which are limited to a maximum of 1200 lines/mm.

5. Each grating is made to order, rather than being selected from a limited catalogue.

6. The DCG layer containing the hologram is protected between a substrate and a cover, both of glass. The exposed surfaces can be anti-reflection coated for the optimum angle and wavelength.

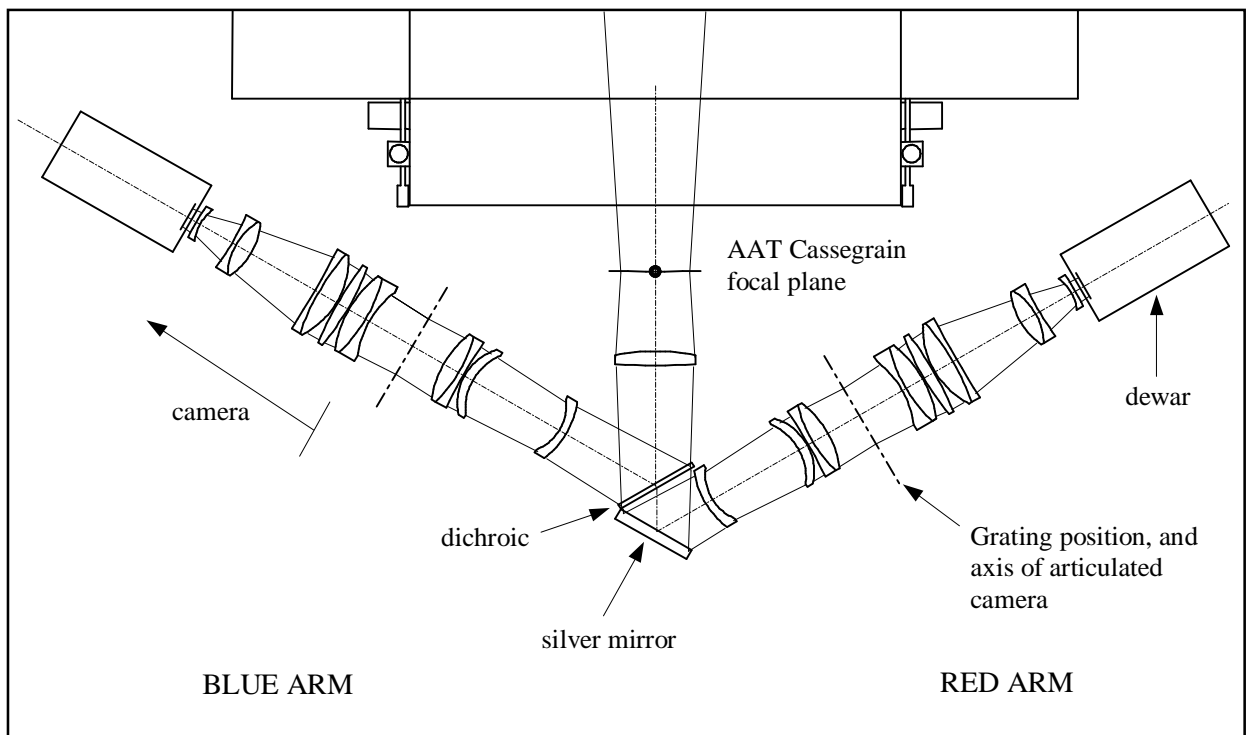

**Figure 1:** Schematic layout of ATLAS in the Cassegrain cage of the AAT, including optics and detector dewars but omitting the enclosure and mechanical mounting. The beam into the blue arm is selected and folded by a dichroic. The beam into the red arm is also folded, to fit the instrument into the available space. In this diagram the motion of the articulated cameras is out of the page. A filter wheel, also incorporating Hartmann shutters, will be included close to the grating in each arm. The optical design for the collimator and camera has been produced by Damien Jones (Prime Optics, Qld, Australia).

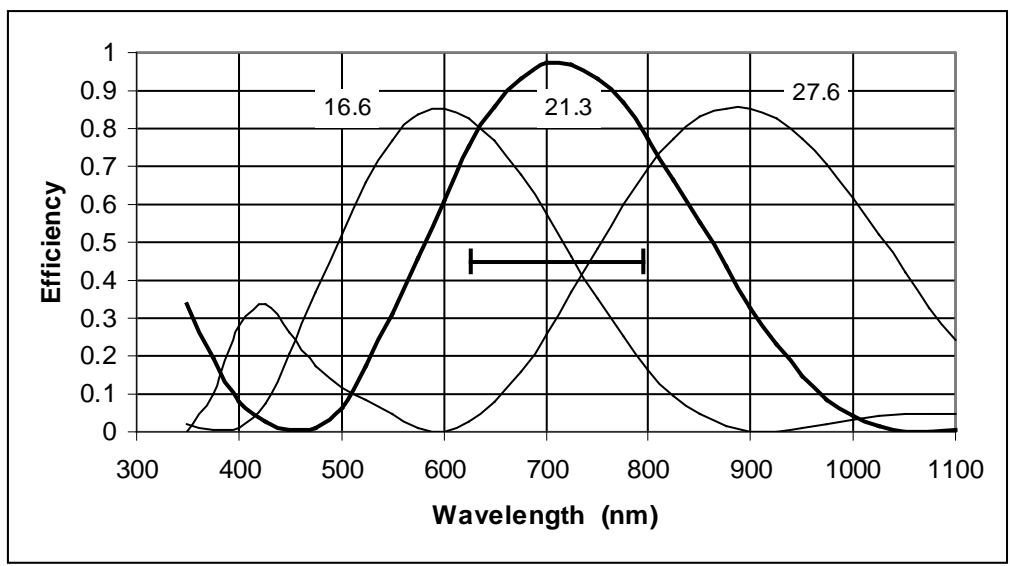

**Figure 2:** Theoretical efficiency vs wavelength curves for a grating of 1020 l/mm, at three different incidence angles (degrees, in air). The calculations used a rigorous coupled-wave simulation, and assumed a DCG layer of thickness 6.5 μm, mean refractive index 1.4, and sinusoidal refractive index modulation to extremes of ±0.057. These parameters give optimum performance at λ712 nm. Losses due to reflections at the surfaces have not been included, but should be only 1-2% with anti-reflection coatings. The horizontal 'scale bar' indicates the range of wavelengths received by the (red arm) detector when centred on λ712 nm, and using an *f*/2.2 camera. For other centre wavelengths the received wavelength range would be altered by resetting the articulated camera.

The problems associated with the use of VPH gratings at the present time are:

1. The response vs wavelength along the detector at any one setting is more strongly peaked than for a surface relief grating. This effect can be diminished by making the DCG layer as thin as possible (consistent with a practical upper limit ~±0.1 to the amplitude of the refractive index modulations in the grating). Nevertheless, a ruled-surface grating system with a suitably large detector array might well achieve a larger useful wavelength range.
2. The extra mechanical complication of the articulated camera is a significant issue. The grating tilt relative to the incoming beam must also be variable.
3. VPH gratings are as yet unproven in a large intermediate-dispersion astronomical spectrograph, although one is already successfully in use for low-dispersion survey spectroscopy at the AAT[3].
4. The availability and price of VPH gratings of the required size and quality are still somewhat uncertain. We are aware of three possible manufacturers, Ralcon Development Laboratory, Kaiser Optical Systems Inc, and Broadbent Holographics, all in USA.
5. In multi-object spectroscopy, the effective blaze wavelength shifts significantly for objects off-axis in the wavelength direction (see Section 5).

# 3. ATLAS DESIGN DECISIONS

This section discusses the major design decisions which have resulted in the present form of the ATLAS concept.

### 4.1 Transmission gratings only

Having accepted the need for an articulated camera, the instrument can be very versatile. At the 'straight through' position (deviation angle 0°) it can do direct imaging; at deviation angles of about 20° it performs low dispersion spectroscopy, and at up to ~90° gives high dispersion spectroscopy. Initially it appeared attractive to allow deviation angles up to ~135°, i.e. a collimator-camera included angle of 45°. This would have allowed use of conventional reflection gratings, further increasing the versatility and providing a backup in case of difficulties in obtaining suitable VPH gratings. But in order to allow an included angle of 45° without the collimator and camera optics colliding, the pupil relief distance between the grating and the nearest element of both the collimator and the camera would have to be substantially increased. This would increase the size and cost of the optics, and lose the advantage of the large field of view. Thus we decided to commit the instrument to the sole use of transmission gratings as dispersive elements, and to obtain the required resolution they must use VPH technology.

### 4.2 Collimated beam diameter

The high line density obtainable with VPH gratings makes it possible in principle to achieve high spectral resolution with even a small collimated beam diameter, which would enable the instrument to fit more easily into the Cassegrain cage, and reduce the cost of both the optics and the mechanical components. The expression for resolution ($R = \lambda/\delta\lambda$) is

$$R = \frac{2 b \tan \theta_i}{D \theta_s} \tag{1}$$

where $b$ is the collimated beam diameter, $\theta_i$ is the incident angle from the grating normal (in air, not in the grating), $D$ is the telescope diameter and $\theta_s$ is the slit width in radians on the sky. This equation applies only for Littrow configuration and when immersion prisms are not used. By using high line density gratings, the beam deviation and hence $\tan\theta_i$ can be increased. But this is achieved at the expense of throughput: at a deviation angle of 60° within the grating medium there is a 50% loss of the p-plane polarization, resulting in a drop to 75% efficiency for unpolarized light, in addition to other losses. At 90° deviation the p-plane light is lost entirely. Increasing the resolution by decreasing $\theta_s$ involves either losses at a narrow slit or losses in an IFU.

In the ATLAS design we have opted for the largest practical collimated beam diameter as defined by space constraints within the Cassegrain cage. The resulting 150 mm beam enables ATLAS to maintain good resolution without sacrificing throughput by using excessive values of the deviation angle. We take a deviation angle of 60° within the grating as the nominal maximum, and ATLAS can achieve a resolution of 9,500 at this point, with a 1.5″ slit. A further advantage of the large beam diameter is that it minimises the problems with blaze shift for off-axis objects (see Section 5).

### 4.3 Dual-beam format

The Advisory Committee on Instrumentation for the AAT recommended that ATLAS should be a planned as a dual-beam instrument from the beginning, rather than leaving a second beam as a possible later addition. A dual-beam configuration will increase observing efficiency for many projects by approximately doubling the wavelength range observed at one time. It will also enable optimised coatings and detectors to be used in each of the blue and red arms.

### 4.4 Camera focal ratio

We have not yet finalised the camera speed. A focal length of 330 mm (nominal $f/2.2$) gives a spatial field of 24′, at 0.36″ per pixel. The collimator and camera designs both perform well over this field. This camera speed gives 4.1 pixel sampling for a 1.5″ slit. A degree of oversampling (i.e. more than 2 pixels per FWHM) always helps the quantitative interpretation of a spectrum. Line profiles are better defined in the presence of noise when moderately oversampled[4,5], and for projects requiring precise wavelengths this is particularly important. If there is any deficiency in charge transfer on the CCD detectors, oversampling helps mitigate its effect. Moreover, while ATLAS is designed for a slit width matched to the median seeing of 1.5″, it is also important that it be usable with narrower slits. At $f/2.2$ we have 2.7-pixel sampling of a 1″ slit, and 2 pixel sampling at 0.75″, the latter suitable as an IFU output and giving R up to 19,000. Thus a camera of $f/2.2$ gives important versatility to ATLAS.

Nevertheless, the wavelength range on each detector could be increased by using a faster camera, and this would increase the observing speed if wide wavelength ranges are needed. Thus we are investigating the possibility of cameras down to a focal length of 255 mm (nominal $f/1.7$). At this focal ratio the angular scale would be 0.468″ per pixel, and the field of view on the detector could be 31.9′. It is not yet clear whether the collimator could be redesigned to handle such a large field, or whether some of the detector would be left unused in the spatial direction. If this were necessary, it would reduce the number of objects which could be observed with a single multi-object mask. The implications for sampling of narrow slits and an IFU must also be considered.

### 4.5 Use of etalons

The ATLAS design is capable of performing direct imaging when the articulated camera is set to the straight-through position. This gives the possibility of carrying out tunable-filter imaging with etalons, as currently done using the Taurus Tunable Filter (TTF)[6]. With ATLAS already intended to result in the decommissioning of the RGO Spectrograph and the Low Dispersion Survey Spectrograph, it would be attractive for operational economy if it could also replace Taurus/TTF. ATLAS will also have a larger field size than Taurus. However, the choice of a large (150 mm) beam size in order to maximise the quality of spectroscopy has virtually precluded operation with full-size etalons because their cost would be prohibitive. Operation over a smaller field could be considered by using the $f/15$ secondary mirror, but this still results in a beam size of 80 mm, which would result in vignetting losses if the present Taurus etalons were used. It remains an option to obtain etalons of diameter ~100 mm, suitable for the 80 mm beam, and accept the reduction in field size to 13′. This still exceeds the current Taurus field of 9′.

## 4. SPECTRAL RESOLUTION

A survey of the AAT user community found that most observers would be satisfied with resolutions up to 5,000 (i.e. 60 km/s), although this expectation was no doubt conditioned to some extent by what had been available with the RGO Spectrograph. However, a significant number wanted *R* up to 10,000. Given the greatly improved throughput of ATLAS compared with older instruments such as the RGO Spectrograph, and the possibility of obtaining high resolutions in first order using VPH gratings, we set a target working resolution for the standard mode of about 10,000. Still higher resolution could be obtained by using narrower slit widths or an IFU (or accepting higher losses of the p-plane polarization).

Figure 3 shows resolutions for three representative configurations. Values have been calculated numerically, by finding the increment in wavelength $\delta\lambda$ which displaces the diffracted beam by the same angle as the shift in input angle due to the projected slit width (i.e. $\delta\lambda$ is the wavelength equivalent of the slit width). In practice, with the object signal peaked somewhat towards the centre of the slit, and minor blurring due to optical aberrations, $\delta\lambda$ will be essentially equal to the FWHM of a monochromatic spectral feature. Equation (1) gives the same result if no immersion prisms are used. Where prisms are used, a generalized form of equation (1) exists[7], but the numerical solution is preferred because it also allows for the dispersion of the prisms. Immersion prisms are needed for the higher dispersion gratings where the angles of incidence and diffraction in air would otherwise be excessive. Since the gratings are used in the symmetrical Littrow mode, equal angle prisms are used on both sides of the grating and effectively act as input and output couplers. Calculated efficiency curves for a 1020 l/mm grating are given in Figure 2.

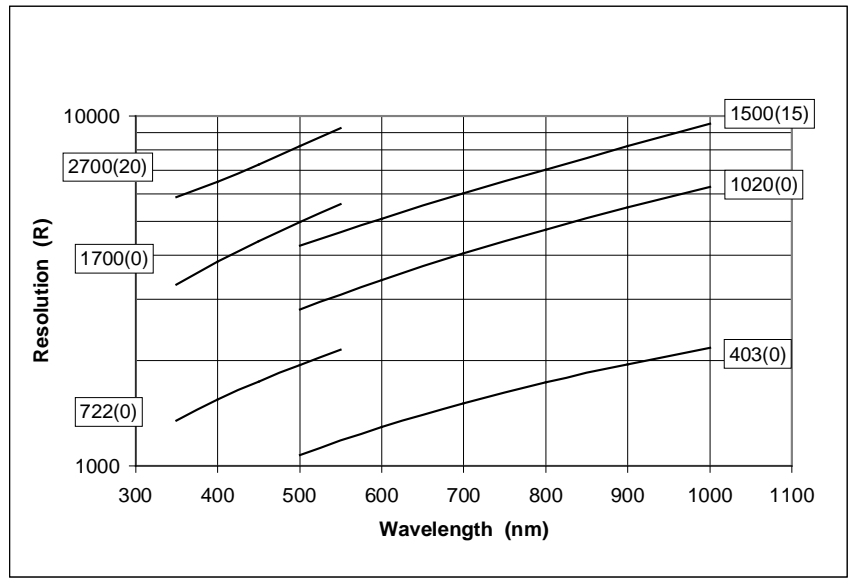

**Figure 3:** Spectral resolution ($R = \lambda/\delta\lambda$) of ATLAS for a representative set of gratings. Each curve is annotated with the line density (lines/mm) and the apex angle of immersion prisms used on both sides of the grating. Zero indicates no prism. A slit width of 1.5″ is assumed. Each point on the curves shows the resolution as if set up with that wavelength as the centre wavelength. Resolutions were calculated numerically, and take account of the prism dispersion (in this case assuming BK7). The curves are cut off at $\lambda$ = 550 nm or 1 μm. The two high dispersion gratings also reach the nominal limit of deviation angle within the grating medium (60°) at these wavelengths. The two low dispersion gratings are together able to cover the entire spectrum in a single setting.

## 5. VPH BLAZE SHIFT FOR OFF-AXIS OBJECTS

VPH gratings are subject to an effect which is not significant for conventional ruled-surface gratings, namely a marked shift in the effective 'blaze' wavelength for objects which are off-axis in the wavelength direction. This is of importance for multi-slit survey observations.

The effect arises from the way the Bragg condition varies with incident angle. As Figure 4 illustrates, the Bragg condition is equivalent to specular reflection from the 'fringe' planes of the grating, which are normal to the grating plane for symmetrical (Littrow) ATLAS gratings. For the 'on-axis object' the grating tilt will be adjusted so that the Bragg 'blaze' peak for the desired centre wavelength $\lambda_C$ is in the same direction as the diffraction of that wavelength, the latter being set by the usual diffraction condition such that the path difference $a + b = \lambda_C / n_2$, (where $n_2$ is the refractive index within the DCG layer). In other words, the articulated camera will be set at the correct angle to receive $\lambda_C$, and the grating tilt will be set to tune the peak efficiency to the same wavelength. An object off axis in the wavelength direction results in the incident light encountering the grating at a different angle, as illustrated in exaggerated form in the lower part of Figure 4. In effect, the grating is differently tilted for this object, and so is tuned to peak efficiency at a different wavelength. The new Bragg blaze direction follows the specular reflection rule, but the desired centre wavelength is displaced in the opposite direction, to maintain the path difference $c + d = \lambda_C / n_2$. Thus some different wavelength is now diffracted in the direction of the Bragg condition, i.e. the 'blaze' wavelength has shifted. (Note that there is no significant effect for objects off axis in the spatial direction: their collimated beams encounter the grating at the same angle to the fringe planes as for an on-axis object.)

The magnitude of the off-axis angles can be quite significant, since field angles on the sky are magnified by the ratio of the telescope focal length to the collimator focal length, or equivalently $D/b$. In ATLAS an object 12′ from the axis will result in a collimated beam making an angle of 5.2° to the on-axis beam. However, objects would not be placed at the edge of the field in the wavelength direction; a limit of ±6′ would be more reasonable, leading to ±2.6° at the grating.

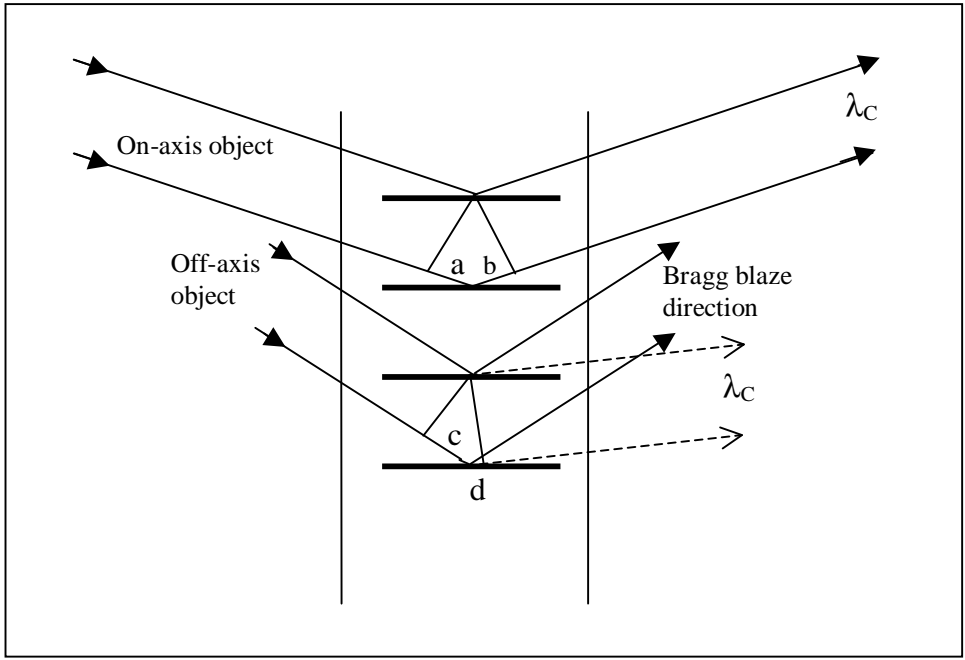

**Figure 4:** Schematic illustration of ray paths through a VPH grating for both an on-axis object and an off-axis object. For simplicity the refraction at the grating surface is not shown. It has been allowed for in the analysis.

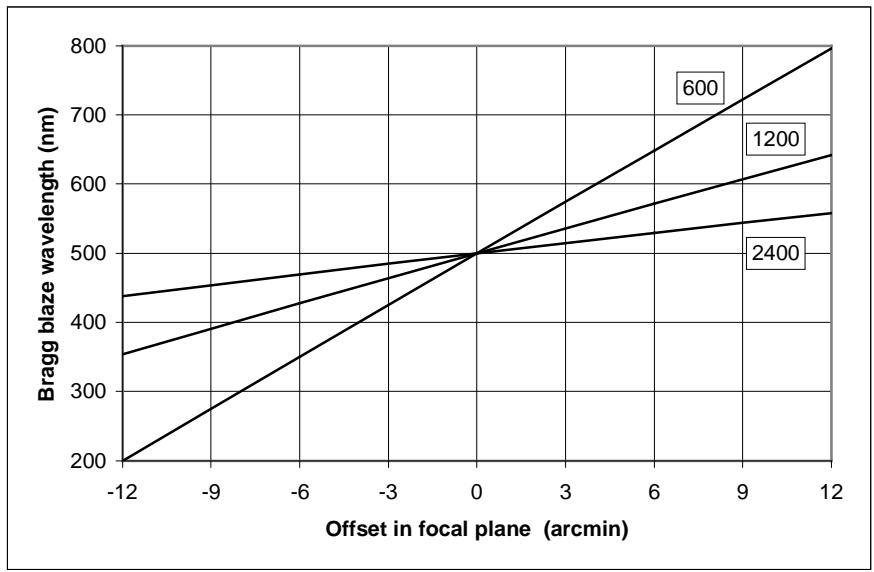

**Figure 5:** Calculated wavelengths of peak efficiency ('blaze' wavelengths) as a function of angle off-axis in the wavelength direction, for the ATLAS spectrograph. An on-axis wavelength of 500 nm is assumed. Relations are shown for gratings of 600, 1200 and 2400 l/mm.

It is simple to calculate the value of wavelength that is diffracted into the same direction as the Bragg blaze condition, and Figure 5 shows the results for three representative grating line densities. Note that one would not want to use more than about ±6′ from the field centre. It can be shown that for the simple case as in Figures 4 and 5 (untilted fringes, no immersion prisms) the slope of the blaze shift is given by the expression

$$\frac{\Delta \lambda}{\Delta \phi} = \frac{\pi}{5400} a \frac{D}{b} \cos \theta_i \quad \text{nm/arcmin} \qquad (2)$$

where $\Delta\lambda$ is the blaze shift in nm, $\Delta\phi$ is the off-axis angle in arcmin on the sky, $a$ is the grating line spacing in nm, $D$ is the telescope diameter, $b$ is the collimated beam diameter (in the same units as $D$) and $\theta_i$ is the air incident angle. The maximum blaze shift will depend on the range of field angles at the grating that are produced by the collimator and accepted by the camera. Despite the large (24′) field of ATLAS, the magnitude of the blaze shift is moderated by the large collimated beam diameter.

Although the blaze shift has the largest magnitude for the low dispersion gratings, a more significant issue is the size of the shift compared with the bandwidth of the grating. In this sense the shifts are largest for high resolution gratings – i.e. at the maximum field offset the desired central wavelength falls further away from the peak of the efficiency curve for high resolution cases. For ATLAS, assuming a grating of thickness 3.7 μm, and using the formula for bandwidth given by Barden *et al.*[2], the ratio of blaze shift at 12′ to HWHM of the efficiency curve is 0.27, 0.53 and 0.97 for the 600, 1200 and 2400 line gratings respectively. At a more practical limit of ±6′ for the field edge none of these shifts would be severe at this grating thickness.

The blaze shift for off-axis objects will have a number of implications: 1) The spectral response will vary with position in the field, complicating the selection and completeness criteria for surveys. 2) It may be necessary to place objects in the slit mask over a smaller range in position in the wavelength direction than would otherwise be the case. 3) Fluxing of spectra will be more complicated, requiring calculation of the shift of the spectrophotometric response for each slit, and/or use of flat field calibration. 4) The use of filters to limit the length of spectra and so fit more objects into one exposure would be less advantageous, since the blaze peak could move out of the filter bandpass.

With suitable design of the spectrograph and the grating, the blaze shift effects can be reduced to an acceptable level, giving efficiencies that remain greater than those of reflection gratings throughout the observed wavelength range. The effect may be regarded as a complication in the design and use of VPH-based spectrographs, but not having significant impact on the overall efficiency.

## 6. MULTI-ORDER VPH GRATINGS

Some preliminary tests of VPH gratings in orders up to $m = 3$ were reported by Barden *et al.*[2], who found surprisingly high efficiencies in the higher orders. Rigorous coupled-wave analysis indicates that non-sinusoidal modulation of the refractive index may be required for good efficiency in higher orders. Future developments in this area could be most worthwhile, because if it becomes possible to produce VPH multi-order 'echellette' gratings, they would make a very useful addition to the capabilities of a general-purpose instrument such as ATLAS. We take the term 'echellette' to refer to gratings used in orders from about 3 to 15. (A better term would be desirable, since 'echelle' or 'echellette' has clear connotations of the 'staircase' profile of an echelle reflection grating.)

The interesting aspect of a VPH echellette is that tilting the grating to a different angle of incidence will tune it in *resolution* rather than in wavelength. A first order VPH grating is tilted (i.e. 'tuned') in order to satisfy the Bragg condition at the same angle at which the desired wavelength will emerge – so placing the blaze efficiency peak on the desired centre wavelength. This amounts to setting the blaze efficiency peak to a certain band of exit angles from the grating. For multi-order gratings, what this means is not selection of a wavelength band within one order, but selection of which orders receive the light at high efficiency. Thus a grating which is used at R ~ 3000 in orders 3 to 9 can also be used at R ~ 5000 in orders 4 to 13. The tuning now selects which orders are used. As with first order gratings there will be a limit to the extent over which this tuning can be varied without excessive loss of efficiency.

Cross dispersion can be provided by prisms. Their mounting would be similar to that of immersion prisms, but with the prisms rotated with respect to the grating lines. A further benefit of the versatility provided by the articulated camera is that the new direction of the cross-dispersed beam can be handled simply by rotating the grating/prism assembly so that the exit beam is in the plane of the articulated camera, and then setting the camera to the required angle. The slit will then need to be rotated to match the new direction of the grating lines. Detector rotation may be desirable but is not essential.

Figure 6 shows the wavelength ranges on the detector for various orders, and the resolution, for a grating at two different angles of incidence. In ATLAS the spectrum would be split into the blue and red arms, with gratings optimised (in DCG thickness and refractive index modulations) separately for the blue and red. Adequate cross dispersion can be obtained in this example with one LF5 prism of apex angle 20° on each side of the gratings, providing order separations of not less than 24″ on the sky.

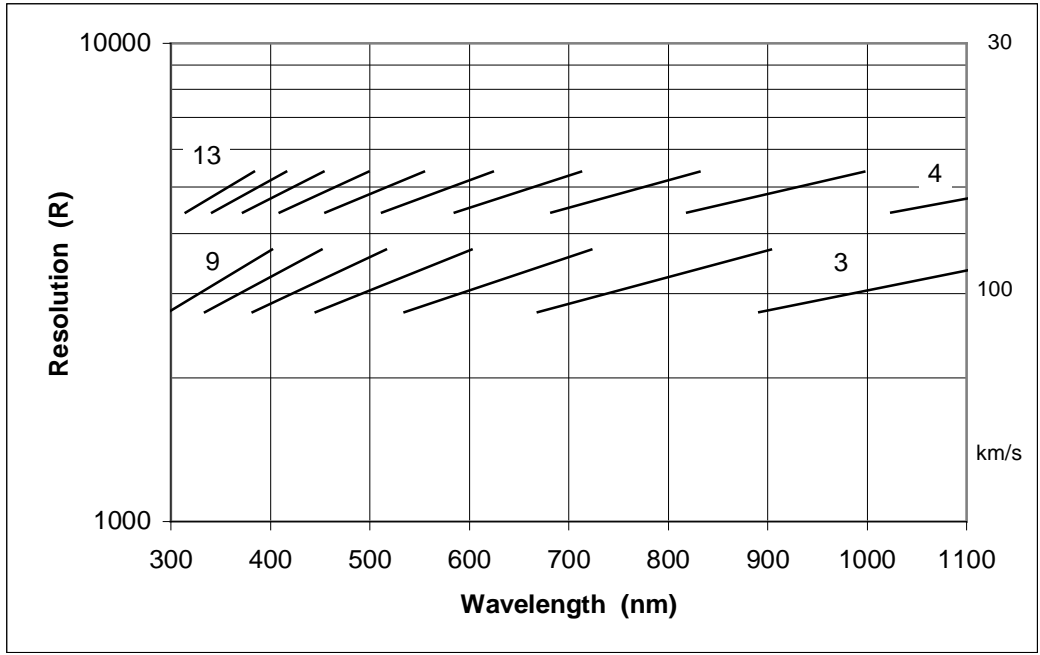

**Figure 6:** Resolution ($\lambda/\delta\lambda$) vs wavelength for a hypothetical multi-order grating with 185 l/mm. Each line segment shows the resolution and wavelength range on the detector for a single spectral order, assuming an *f*/2.2 camera. The bottom row is for an incident angle to the grating (in air) of 16.9°, while the top row is for an incident angle of 24.9°. The numerical annotations show order numbers. Resolution up to a maximum of about 9,000 could be obtained by using still higher orders, at a larger tilt angle. However, a lower line density grating would then be preferred in order to avoid gaps in the wavelength coverage, at least for the red arm.

## 7.  PRESENT STATUS

There are a number of significant engineering challenges to be met for successful completion of the ATLAS design, and these are now under study. Foremost among them is the design for the articulated camera, which must hold the image on the detector to high stability, but be adjustable over a range of at least 90°. Our target is that image motion due to flexure and/or thermal effects should not exceed 0.1 pixel per hour of observation. It is likely that active compensation will be required to achieve this level of stability[8]. Compensation cannot be achieved by moving the grating (which simply re-tunes the broad peak of efficiency) but instead it will be necessary to move the detector. We aim to achieve a design with sufficiently small hysteresis that a look-up table can supply the necessary data for the active compensation, rather than requiring a closed-loop system with flexure measurement.

The ATLAS project is aiming to reach the Concept Design Review stage during 2000.